NONLINEAR DYNAMICS

# Revealing physical interaction networks from statistics of collective dynamics

Mor Nitzan,[1,2,3] Jose Casadiego,[4,5] Marc Timme[4,5,6,7,8]*

Revealing physical interactions in complex systems from observed collective dynamics constitutes a fundamental inverse problem in science. Current reconstruction methods require access to a system's model or dynamical data at a level of detail often not available. We exploit changes in invariant measures, in particular distributions of sampled states of the system in response to driving signals, and use compressed sensing to reveal physical interaction networks. Dynamical observations following driving suffice to infer physical connectivity even if they are temporally disordered, are acquired at large sampling intervals, and stem from different experiments. Testing various nonlinear dynamic processes emerging on artificial and real network topologies indicates high reconstruction quality for existence as well as type of interactions. These results advance our ability to reveal physical interaction networks in complex synthetic and natural systems.

## INTRODUCTION

Many complex systems in physics and biology constitute networks of dynamically interacting units (1). Examples range from gene regulatory networks in the cell (2–5) and neural circuits in the brain (6–8) to food webs in ecosystems (9) and power grids (10–13), as well as other supply systems of engineering (14, 15). These systems' interaction networks fundamentally underlie their collective dynamics and function, thus rendering the knowledge of their interaction topology essential. For instance, identifying new pathways in gene regulatory networks and understanding long-range feedback in engineering systems require exact knowledge of their physical interaction networks.

A fundamental question about both natural and artificial networks is thus which units directly act upon which other units. For instance, for networks of interacting units $i \in \{1, ..., N\}$ described by ordinary nonlinear differential equations $dx_i/dt = F_i(\mathbf{x})$, this question mathematically becomes which of the variables $x_j$ among $(x_1, ..., x_N) =: \mathbf{x}$ explicitly appear in $F_i(\mathbf{x})$. In many settings, this physical connectivity is not directly accessible. Revealing these interaction networks then poses a high-dimensional nonlinear reconstruction problem. Evaluating the dynamics of the system exploiting heuristic approximations, such as thresholding correlations or other measures of statistical dependency between units' dynamics, is simple, efficient, and generally feasible, yet often an unreliable predictor for physical connectivity (16). A conceptual reason is that more than one statistical dependency network may emerge for the same physical network, for example, due to multistability (17–19). In turn, methods aiming at directly identifying physical interactions generally require either a priori knowledge of a detailed model of the system, rely on the system being in simple states (such as close to fixed points), or need high-resolution, synchronized, temporally ordered observations for all units with connections of interest (20). For example, transfer entropy (21) and cross-embedding (22) require temporally ordered measurements; a direct method for inferring structural connectivity described by Shandilya and Timme (23) requires synchronous, temporally ordered, high-resolution measurements and prior knowledge of the system's model; and the system identification method described by Gardner et al. (5) requires the system to be at steady state and the dynamics to be essentially linear in the activity of the nodes of the network. However, real systems, such as genetic regulatory networks, other biological circuits, and even some human-made systems (24), often do not fulfill these requirements. To the best of our knowledge, a method capable of inferring physical interaction networks without requiring at least one of these constraints does not exist to date.

Here, we propose a generic strategy to reveal physical interactions from responses of invariant measures (that is, distributions of points sampled in state space) to small driving signals. The strategy does not rely on any of the requirements above. Via compressed sensing, the resulting equations obtained from driving-response experiments yield the network structure even if the number of available experiments is small compared to the network size. Because only statistics of recorded system states are evaluated to reveal the physical connectivity, measurements of dynamic states can be temporally disordered, be acquired at large sampling intervals, come from different experiments, and be collected at different times for different units. In addition, no detailed prior knowledge for the model of the system is required.

## THEORY BASED ON TIME INVARIANTS

To introduce the basic strategy of reconstructing networks from time invariants (Fig. 1), we consider networks of units $i \in \{1, ..., N\}$ represented by state variables $x_i(t) \in \mathbb{R}$ evolving in time $t$ according to

$$\dot{x}_i(t) = F_i(x) + \xi_i(t) \quad (1)$$

Here, $x(t) = (x_1(t), ..., x_N(t)) \in \mathbb{R}^N$ is the state vector of the entire network, $\dot{x}_i(t) \equiv \frac{d}{dt} x_i(t)$ denotes the temporal derivative of the variable $x_i(t)$, and $\xi_i(t)$ represents noise with zero average. Systems of higher dimensional units, $x_i(t) \in \mathbb{R}^d$, $d > 1$, are discussed further below and in note S1.

Driving the system (Eq. 1) with signals $I_i^{(m)}(t)$

$$\dot{x}_i^{(m)} = F_i(x^{(m)}) + \xi_i^{(m)} + I_i^{(m)} \quad (2)$$

[1]Racah Institute of Physics, Hebrew University of Jerusalem, 9190401 Jerusalem, Israel. [2]Faculty of Medicine, Hebrew University of Jerusalem, 9112001 Jerusalem, Israel. [3]School of Computer Science, Hebrew University of Jerusalem, 9190401 Jerusalem, Israel. [4]Network Dynamics, Max Planck Institute for Dynamics and Self-Organization, 37077 Göttingen, Germany. [5]International Max Planck Research School for Physics of Biological and Complex Systems, 37077 Göttingen, Germany. [6]Technical University of Dresden, Institute for Theoretical Physics, 01062 Dresden, Germany. [7]Bernstein Center for Computational Neuroscience, 37077 Göttingen, Germany. [8]Department of Physics, Technical University of Darmstadt, 64289 Darmstadt, Germany.
*Corresponding author. Email: timme@nld.ds.mpg.de

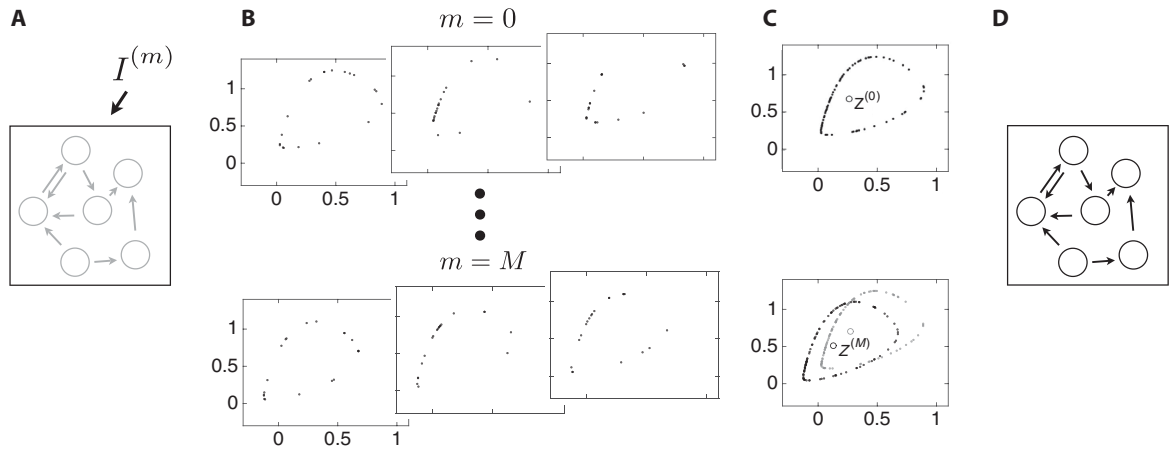

**Fig. 1. Strategy of network reconstruction from responses of time invariants.** (**A**) A networked dynamical system with unknown topology (gray) is perturbed by external driving signals $I^m$, $m \in \{1, …, M\}$. (**B**) Potentially noisy, disordered, low-resolution data are collected from several different experiments. (**C**) The centers of mass $z^{(m)}$ of each of these distributions of points sampled in state space are calculated. (**D**) The changes $z_i^{(m)} - z_i^{(0)}$ in response to driving signals $I^{(m)}$ yield the network topology.

modifies its dynamics. For temporally constant or otherwise stationary driving signals, the temporal trajectories of the system after potential transients may exhibit collective dynamics that generate a defined statistics of points in state space. These states are recorded under $M$ different driving conditions $m \in \{1, …, M\}$ with signals $I_i^{(m)}$ and, for each experiment, generate an invariant density $\rho^{(m)}$ characteristic of the dynamics defined by Eq. 2. One simple quantity induced by such a density is given by its center of mass $z^{(m)} = <x^{(m)}>_{\rho^{(m)}}$. If these data sample the invariant measure well (25), then the average may be approximated by the sample mean $z^{(m)} = <x^{(m)}(t)>_{t \in T} = |T|^{-1} \Sigma_{t \in T} x^{(m)}(t)$ (see note S1 for a more detailed discussion). Here, $T$ is the set of time points at which the data are recorded.

How can we reconstruct interaction networks from these data? Approximating Eq. 2 up to first order around $z^{(0)}$ yields

$$\dot{x}_i^{(m)} \approx F_i\left(z^{(0)}\right) + \sum_{j=1}^{N} \frac{\partial F_i}{\partial x_j}\bigg|_{z^{(0)}} \left(x_j^{(m)} - z_j^{(0)}\right) + \xi_i^{(m)} + I_i^{(m)} \quad (3)$$

The exact conditions under which this approximation is justified are elaborated in note S1. By averaging over the recorded sample points $T$, we get

$$\dot{z}_i^{(m)} \approx F_i\left(z^{(0)}\right) + \sum_{j=1}^{N} \frac{\partial F_i}{\partial x_j}\bigg|_{z^{(0)}} \left(z_j^{(m)} - z_j^{(0)}\right) + \bar{I}_i^{(m)} \quad (4)$$

where $\bar{I}_i^{(m)} := <I_i^{(m)}(t)>_{t \in T}$ are temporal averages of the stationary driving signals, and we set $I_i^{(0)} \equiv 0$ for all $i$. Last, substituting the expression for the undriven dynamics $F_i(z^{(0)})$ by setting $m = 0$ in Eq. 4, we obtain (see note S1 for a detailed derivation)

$$\dot{z}_i^{(m)} - \dot{z}_i^{(0)} \approx \sum_{j=1}^{N} J_{ij}\left(z_j^{(m)} - z_j^{(0)}\right) + \bar{I}_i^{(m)} \quad (5)$$

where $J_{ij} := \frac{\partial F_i}{\partial x_j}\big|_{z^{(0)}}$ are the elements of the Jacobian $J = DF|_{z^{(0)}}$. We take $\dot{z}_i^{(m)} = \dot{z}_i^{(0)} = 0$, because the centers of mass do not change in time if the recorded points sample the invariant density well (see note S2 for error estimates for sampling). This yields

$$-\bar{I}_i \approx \Delta z \, J_i^\mathsf{T} \quad (6)$$

where $\bar{I}_i \in \mathbb{R}^{M \times 1}$ is the vector of averaged driving signals $\bar{I}_i^{(m)}$, $\Delta z \in \mathbb{R}^{M \times N}$ is the matrix of response differences $z_j^{(m)} - z_j^{(0)}$ of the centers of mass, and $J_i \in \mathbb{R}^{1 \times N}$ is the respective row of the Jacobian matrix.

Evidently, the differences in the invariant density's centers of mass are jointly determined by the driving vector and the network topology. We remark that the reconstruction problem decomposes over nodes in the network such that the set of incoming interactions to each node can be reconstructed independently. A sufficient number of driving-response experiments thus yield a set of linear equations (Eq. 6) for each node $i$, restricting the potential interaction networks estimated by $J$. Our goal of identifying which variables $x_j$ appear in $F_i(\mathbf{x})$ is thus equivalent to finding those pairs $(i, j)$ where $J_{ij} \neq 0$ such that unit $j$ directly acts on $i$ (and thus also those where no such direct interactions exist, $J_{ij} = 0$). Notice that, because the Jacobian is evaluated at the center of mass of the unperturbed invariant density, the reconstruction approach is expected to recover the correct interactions if they consistently exist across the relevant fractions of state space, which include the observed driven dynamics and the unperturbed centers of mass. Here, we consider constant driving signals, $I_i^{(m)}(t) = \bar{I}_i^{(m)}$, complemented by additive noise. If the number $M$ of experiments is larger than the network size $N$, then reconstructing a given network becomes possible via a least squares solution to Eq. 6. However, in many experimental settings, the number of available experiments is relatively small. To overcome this limitation, we exploit compressed sensing theory by determining an $L_1$-norm minimum solution $\hat{J}_i$ to Eq. 6 such that the number of experiments $M$ can be much smaller than the network size $N$ [see Methods; (26–29)]. Last, we rank the resulting absolute values $|\hat{J}_{ij}|$ and vary a threshold $J_c$ to distinguish between existing ($|\hat{J}_{ij}| \geq J_c$) and absent ($|\hat{J}_{ij}| < J_c$) interactions. Hence, evaluation of inference performance is done in a parameter-free manner (note S3).

## PERFORMANCE ON ARTIFICIAL AND REAL NETWORK TOPOLOGIES

Our strategy is successful in reconstructing the topology of physical interaction networks across a range of systems and collective dynamics (Figs. 2 to 5). To evaluate the quality of reconstruction, we initially consider a class of random networks where each unit is a Goodwin oscillator (30), a prototypical biological oscillator that characterizes various biological processes from circadian clocks to somitogenesis (see Methods for model description) (31–33). Under mild constraints, the strategy readily generalizes to systems of other, also higher-dimensional units (see below) and more complex dynamics (see note S1 for a complete derivation).

We first analyze one network realization (Fig. 2A) and collect individual results to network ensembles to obtain a robust quality analysis. Reconstruction quality is quantified by the area under the receiver operating characteristic (ROC) curve (AUC), a score typically between one-half (chance level) and one (perfect ranking of the reconstructed interactions) (see note S3). As expected, the reconstruction quality improves with the number of driving-response experiments (Fig. 2B).

With compressed sensing methods [see (34–38) and Methods], the number of experiments required to obtain a given quality increases sublinearly with network size and continuously changes with increasing noise (Fig. 2C). Furthermore, the sparser a given network is, the lower the number of required experiments for robust reconstruction (see note S4). The invariants-based reconstruction strategy works analogously for different network topologies and different nodal coupling strengths, both homogeneous and heterogeneous (note S5) and for various dynamic processes; beyond networks of noisy oscillators, robust reconstruction is achieved for networks exhibiting qualitatively distinct dynamics, such as a biological network of transcription factor regulators close to a fixed point and a network of Rössler oscillators exhibiting chaotic dynamics (notes S6 and S7). Finally, the invariants-based strategy outperforms available standard reconstruction baselines, including measures of mutual information and correlation between the activity patterns of every two nodes in the network, partial correlation between the pairwise activity patterns (given the activity patterns of the other nodes in the network), and transfer entropy (note S8 and figs. S7 and S8).

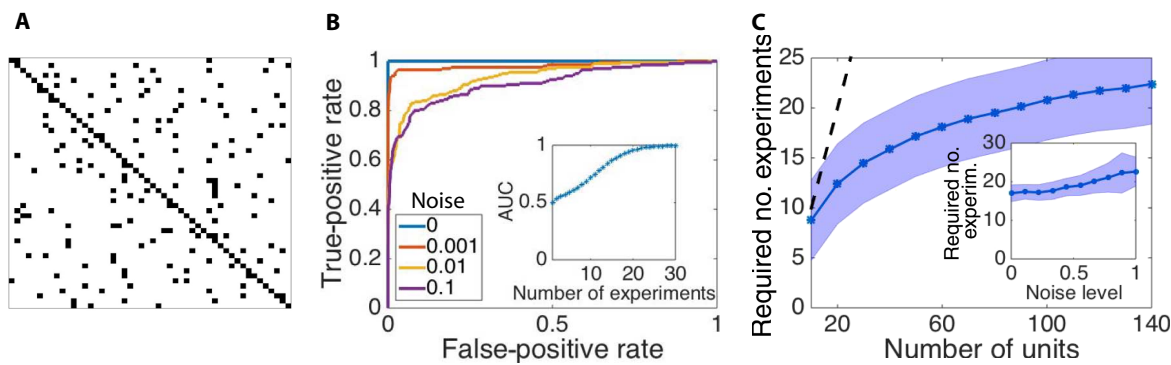

**Fig. 2. Evaluation scheme illustrating robust reconstruction.** (**A**) Representative adjacency matrix indicating network connectivity as defined by present (black) and absent (white) links. (**B**) ROC curve obtained by varying a threshold $J_c$ separating links classified as existing ($|\hat{J}_{ij}| \geq J_c$) from those classified as absent ($|\hat{J}_{ij}| < J_c$) (see note S3). The AUC increases with decreasing noise level, with perfect ranking of reconstructed links in the limit of noiseless dynamics. Inset: The quality of network reconstruction, as specified by the AUC, increases with the number of driving-response experiments. (**C**) The number of experiments required for high-quality reconstruction (here, AUC > 0.95) increases sublinearly (compared to the dotted line) with network size and (inset) changes only weakly with the noise level. Data are shown for random networks of (default size) $N = 50$ Goodwin oscillators with a regular incoming degree of 4, a default noise level of 0.5, a default number of experiments of 25, and a number of sampled time points of 100; shading indicates SD across ensembles of network realizations.

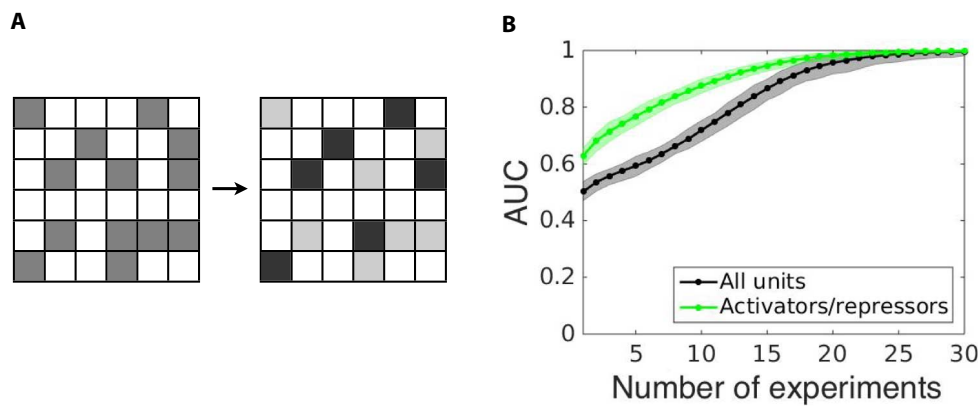

**Fig. 3. Revealing interaction types.** (**A**) Beyond distinguishing existing from missing interactions (schematically represented by the medium gray and white adjacency matrix), activating and inhibiting interactions may be separately detected (dark gray, light gray, and white matrix). (**B**) The reconstruction quality (AUC) benefits from the separate reconstruction of different types of interactions (green curve) compared to joint reconstruction of existing and missing interactions (gray curve), and increases with the number of driving-response experiments. Data are shown for random networks of $N = 50$ Goodwin oscillators with a regular incoming degree of 4 and a number of sampled time points of 100; shading indicates SD across ensembles of network realizations.

In addition to disentangling existing and missing interactions within a network, the strategy can be extended to reveal finer topological features and to require only partial, lower dimensional dynamical information. First, separately ranking the reconstructed $J_{ij}$ values as deduced from Eq. 6 to find activating interactions above a certain threshold and negative ones below a second threshold yields two sets, one for consistently activating and one for consistently inhibiting interactions (Fig. 3). The separate identification of different types of interactions not only provides more information about their nature but also enhances the quality of reconstruction (Fig. 3 and note S9). Second, partial dynamical information, such as experimental data for networks of high-dimensional units limited to only one out of several dynamical variables for each unit, may suffice to reconstruct complete interaction networks (Fig. 4). See note S10 for a systematic evaluation of the effect of missing dynamical information on the quality of reconstruction. These results establish that reconstruction of network interactions is possible in a generic class of dynamical systems under a broad range of conditions.

Beyond generic model systems, the reconstruction strategy also yields promising results for specific biological settings. We demonstrate this on a genetic regulatory network characterizing the circadian clock in *Drosophila* (see note S11 for model description) (*39*). This clock coordinates the biological response to the day-night cycle. The quality of reconstruction of the circadian clock benefits from increasing the number of driving-response experiments and is robust to noisy signals (Fig. 5).

## DISCUSSION

The presented theory relating responses of invariant measures and resulting observables to small driving signals enables us to reveal physical interaction networks from statistics of essentially arbitrary dynamical data that are sufficiently broadly sampled. The theory relies on the system being stationary or otherwise exhibiting an invariant measure that is sampled well by the data. It also relies on the option to drive the system with signals that are sufficiently small to still yield linear responses and, at the same time, sufficiently large to outweigh the noise and finite sampling influence. However, no fine-tuning of the driving signals is needed (see note S12). In addition, the method relies on changes in the stationary dynamics of the system following perturbations. Hence, it suits systems that do not exhibit an extreme form of perfect saturation or systems that adapt on fast time scales in response to external signals. If adaptation time scales are slow relative to the period during which the dynamic sampling is performed, then the method we suggest naturally becomes applicable again, because the system does not fully adapt and is effectively stationary during the measurement time. Thus, shifts in averages can be naturally observed following the same framework. The strategy is successful for generic mathematical model schemes and specific biological settings. Extended options include distinguishing between activating and inhibiting interactions as well as revealing the existence of interactions even if access is limited to only one out of several dynamical variables for individual units.

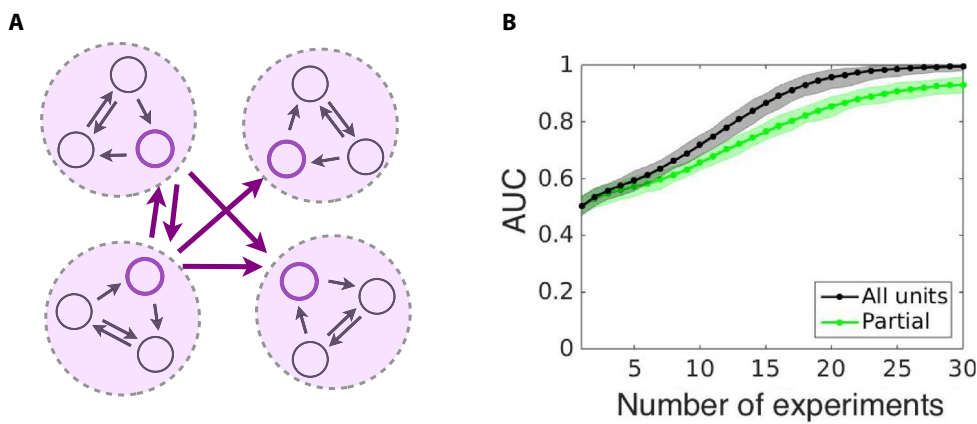

**Fig. 4. Robust reconstruction from one-dimensional sampling of multidimensional unit dynamics.** (**A**) Scheme illustrates three-dimensional units (encircled), coupled through one observed variable (colored), whereas the other two variables are unobserved (gray). (**B**) The reconstruction quality (AUC) stays robust and reduces only slightly for reconstruction based on partial, one-dimensional measurements (green curve) relative to reconstruction based on three-dimensional measurements (gray curve), and increases with the number of driving-response experiments. Data are shown for random networks of $N = 50$ Goodwin oscillators with a regular incoming degree of 4 and a number of sampled time points of 100; shading indicates SD across ensembles of network realizations.

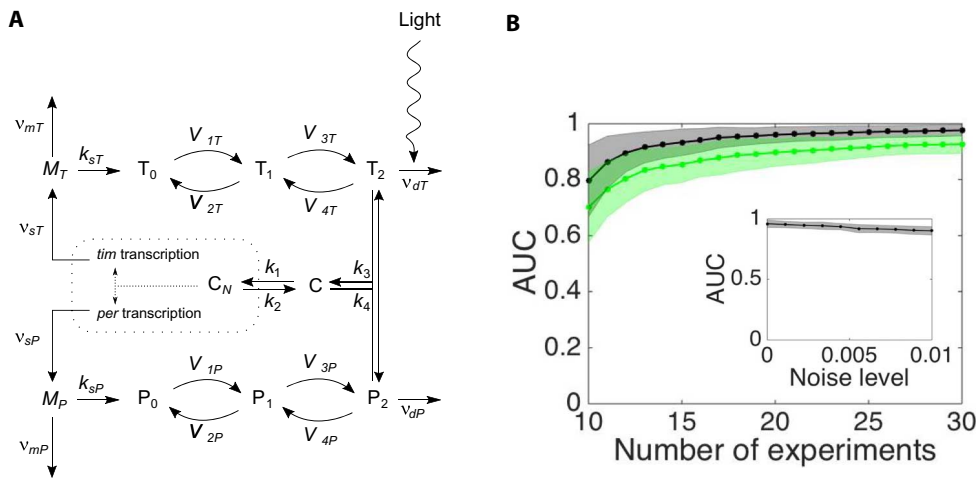

**Fig. 5. Reconstruction of the circadian clock network in *Drosophila*.** The quality of reconstruction (AUC) of the circadian clock (**A**) increases with the number of driving-response experiments, for both noiseless (gray curve) and noisy (green curve) dynamics (**B**), and changes only weakly with noise level (inset). Number of sampled time points, 300; default number of experiments, 20; noise level for noisy case, 0.01; shading indicates SD across ensembles of network realizations. (A) Modified with permission from Leloup and Goldbeter (*39*) (Fig. 1).

Because the strategy relies solely on the invariant density, state space points recorded may be used without known temporal order, may have been recorded at varying sampling intervals, and may even come from several experiments (performed under the same conditions). Furthermore, different units of the same system may, in principle, be recorded separately and at different times. All these options emerge because only statistical information (a rough estimate of an observable, such as center of mass) of the dynamical data is used. This strategy for revealing physical interactions from statistics thus constitutes a previously unknown intermediate approach between purely statistical methods for inferring effective connectivity (for example, correlations) and approaches inferring physical connectivity from high-resolution, time-ordered recordings of the full dynamics (5, 20, 27, 40–42).

The full range of features may be useful under various experimental conditions. For instance, many high-quality measurements of single-cell gene expression levels obtained simultaneously are available either at a system-wide level for a single time point [for example, Taniguchi et al. (43)] or for a few genes for several time points (44), yet there is no restriction in principle to measuring all genes in a sequence of different experiments. Together, these results offer a novel perspective on inferring physical and not only correlative (effective) connectivity of networked systems from statistically sampling their dynamics.

## METHODS

### Goodwin oscillators
To evaluate the quality of the reconstruction approach in a controlled setting, we considered networks of $N$ prototypical Goodwin oscillators $i \in \{1, ..., N\}$ (30), each with three variables—$x_i$, $y_i$, and $z_i$—evolving in time according to $\dot{x}_i = f(z_i) - a_i z_i$, $\dot{y}_i = x_i - b_i y_i - \sum_{j=1}^{N} J_{ij} g(y_i, y_j)$, and $\dot{z}_i = y_i - c_i z_i$. In addition, $f(z_i) = v_o/[1 + (z_i/K_m)^n]$ constitutes a local nonlinearity and $g(y_i, y_j) = y_i - y_j$ constitutes the diffusive interactions. In direct numerical simulations, the parameters are $a_i = b_i = c_i = 0.4$, $v_o = K_m = 1$, and $n = 17$.

### Network topologies
For generic evaluations (Figs. 2 to 4), we used random networks of $N = 50$ Goodwin oscillators with a regular incoming degree of 4. We fixed the degree for Figs. 2 to 4 and varied it to compare networks of different degrees in fig. S2. In addition, we used an Erdős Rényi random network of $N = 50$ genetic regulators, with edge probability $p = 0.01$, where a genetic regulator is chosen to be either an activator or a repressor with equal probability (fig. S3). For the topology of a real biological system, we considered the circadian clock network of Drosophila [(39, 45); see note S11] (Fig. 5).

### Compressed sensing
The framework of compressed sensing (34–38) enables us to reconstruct a high-dimensional sparse signal based on linear measurements, where the number of measurements is small relative to the dimension of the signal. In our context, the goal was to reconstruct the network physical connections $J_i \in \mathbb{R}^{1 \times N}$ for a given unit $i$, by solving the linear set of equations $-\bar{I}_i = \Delta z \, J_i^T$ (Eq. 6) for averaged driving signals $\bar{I}_i \in \mathbb{R}^{M \times 1}$, driving-response matrix $\Delta z \in \mathbb{R}^{M \times N}$, and $M \ll N$. Given that $J_i$ is sufficiently sparse and $\Delta z$ fulfills certain conditions, as elaborated in note S13, this problem can be posed as an $L^1$-norm convex optimization problem with guarantees for a robust and stable solution, and solved using standard software such as CVX, a MATLAB package for specifying and solving convex problems (46, 47).

### SUPPLEMENTARY MATERIALS
Supplementary material for this article is available at http://advances.sciencemag.org/cgi/content/full/3/2/e1600396/DC1
note S1. Details of the derivation of invariant-based reconstruction.
note S2. Error estimates for observables from sampled invariant density.
note S3. Reconstruction evaluation.
note S4. Moderate influence of link density.
note S5. Reconstructing homogeneous and heterogeneous networks.
note S6. Reconstruction of systems near fixed points.
note S7. Reconstruction of chaotic systems.
note S8. Performance compared with available standard baselines.
note S9. Distinguishing activating from inhibiting interactions.
note S10. The effect of missing information.
note S11. Model descriptions.
note S12. The effect of various driving conditions on reconstruction quality.
note S13. Compressed sensing.
fig. S1. Approximating the center of mass of invariant densities by the sample mean.
fig. S2. Sparser networks require fewer experiments for robust reconstruction.
fig. S3. Reconstruction is robust across network topologies.
fig. S4. The quality of reconstruction increases with the number of experiments for a network of genetic regulators.
fig. S5. Reconstruction of a network of Rössler oscillators exhibiting chaotic dynamics.
fig. S6. Comparison of reconstruction quality across different approaches.
fig. S7. Comparison of reconstruction quality against transfer entropy.
fig. S8. Separate reconstruction of activating and inhibiting interactions enhances the quality of reconstruction.
fig. S9. Quality of reconstruction (AUC score) decreases gradually with the fraction of hidden units in the network.
fig. S10. Quality of reconstruction increases as driving signals overcome noise and finite sampling effects.
References (48–50)

**Acknowledgments:** We thank I. Kanter, C. Kirst, B. Lünsmann, and M. Peer for valuable discussions and L. J. Deutsch for help with figure preparation. M.N. is grateful to the Azrieli Foundation for the award of an Azrieli Fellowship. **Funding:** This study was supported by the Federal Ministry of Education and Research (BMBF; grant 03SF0472E) and the Max Planck Society (to M.T.). **Author contributions:** All authors conceived the research and contributed materials and analysis tools. M.N. and M.T. designed the research. All authors worked out the theory. M.N. and J.C. developed the algorithms and carried out the numerical experiments. All authors analyzed the data, interpreted the results, and wrote the manuscript. **Competing interests:** The authors declare that they have no competing interests. **Data and materials availability:** All data needed to evaluate the conclusions in the paper are present in the paper and/or the Supplementary Materials. Additional data related to this paper may be requested from the authors.

Submitted 27 February 2016
Accepted 10 December 2016
Published 10 February 2017
10.1126/sciadv.1600396

**Citation:** M. Nitzan, J. Casadiego, M. Timme, Revealing physical interaction networks from statistics of collective dynamics. *Sci. Adv.* **3**, e1600396 (2017).